\documentclass[twocolumn,showpacs,amsmath,amssymb,aps, prl,superscriptaddress]{revtex4-1}

\usepackage[]{cleveref}
\usepackage[]{graphicx}
\usepackage[]{verbatim}
\usepackage[]{color}
\usepackage{overpic}
\usepackage{rotating}
\usepackage{mathtools}
\usepackage{tabularx}

\makeatletter
\newsavebox{\@brx}

\newcommand{\llangle}[1][]{\savebox{\@brx}{\(\m@th{#1\langle}\)}%
  \mathopen{\copy\@brx\mkern2mu\kern-0.8\wd\@brx\usebox{\@brx}}}
\newcommand{\rrangle}[1][]{\savebox{\@brx}{\(\m@th{#1\rangle}\)}%
  \mathclose{\copy\@brx\mkern2mu\kern-0.8\wd\@brx\usebox{\@brx}}}

  \newcommand{\lllangle}[1][]{\savebox{\@brx}{\(\m@th{#1\langle}\)}%
  \mathopen{\copy\@brx\copy\@brx\mkern4mu\kern-0.7\wd\@brx\usebox{\@brx}}}
\newcommand{\rrrangle}[1][]{\savebox{\@brx}{\(\m@th{#1\rangle}\)}%
  \mathclose{\copy\@brx\copy\@brx\mkern4mu\kern-0.7\wd\@brx\usebox{\@brx}}}

\graphicspath{{./}{./}}

\begin{document}
\title{Kitaev magnetism in honeycomb RuCl$_3$ with intermediate spin-orbit coupling}
\author{Heung-Sik Kim}
\affiliation{Department of Physics and Center for Quantum Materials , University of Toronto, 60 St.~George St., Toronto, Ontario, M5S 1A7, Canada}
\author{Vijay Shankar V.}
\affiliation{Department of Physics and Center for Quantum Materials , University of Toronto, 60 St.~George St., Toronto, Ontario, M5S 1A7, Canada}
\author{Andrei Catuneanu}
\affiliation{Department of Physics and Center for Quantum Materials , University of Toronto, 60 St.~George St., Toronto, Ontario, M5S 1A7, Canada}
\author{Hae-Young Kee}
\email{hykee@physics.utoronto.ca}
\affiliation{Department of Physics and Center for Quantum Materials , University of Toronto, 60 St.~George St., Toronto, Ontario, M5S 1A7, Canada}
\affiliation{Canadian Institute for Advanced Research, Toronto, Ontario, M5G 1Z8, Canada}

\begin{abstract}
Intensive studies of the interplay between spin-orbit coupling (SOC) and electronic correlations in transition metal compounds have recently been undertaken.
In particular, $j_{\rm eff}$ = 1/2 bands on a honeycomb lattice provide a pathway to realize Kitaev's exactly solvable spin model. 
However, since current wisdom requires strong atomic SOC to make $j_{\rm eff}=1/2$ bands, studies have been limited to iridium oxides. 
Contrary to this expectation, we demonstrate how Kitaev interactions arise in 4$d$-orbital honeycomb $\alpha$-RuCl$_3$, despite having significantly weaker SOC than the iridium oxides, via assistance from electron correlations. 
A strong coupling spin model for these correlation-assisted $j_{\rm eff}$ = 1/2 bands is derived, in which large antiferromagnetic Kitaev interactions emerge along with ferromagnetic Heisenberg interactions. Our analyses suggest that the ground state is a zigzag-ordered phase lying close to the antiferromagnetic Kitaev spin liquid. 
Experimental implications for angle resolved photoemission spectroscopy, neutron scattering, 
and optical conductivities are discussed.
\end{abstract}
\maketitle

{\it Introduction} -- Elucidating the cornucopia of novel physical phenomena exhibited by transition metal compounds with electrons occupying $d$ orbitals has been a key focus of modern condensed matter physics. Relativistic effects such as spin-orbit coupling (SOC), which entangles the spin and orbital degrees of freedom, were largely ignored until recently when it was realized that these effects in cohort with electronic correlations could give rise to new ground states, including those with uncommon magnetic ordering.\cite{witczak2014correlated,wan2011topological,Moon_Lab,TIstar,pesin2010mott,lawler2008SL,CK_arxiv,CLK_arxiv} 

In particular, these effects bring about anisotropic exchange interactions that have been suggested as a way to engineer the exactly solvable Kitaev spin model\cite{kitaev2006anyons}  in the honeycomb iridate Na$_2$IrO$_3$. These anisotropic interactions arise between two neighboring iridium (Ir) sites, each with a single $j_{\rm eff}=1/2$ state, through superexchange mediated by the $p$-orbitals on the intervening oxygen atoms that make up the edge-sharing octahedra around each Ir atom.\cite{khaliullin2005orbital,jackeli2009mott} This $j_{\rm eff}=1/2$ state, composed of an equal mixture of $t_{\rm 2g}$ orbitals, which manifests at large SOC $\lambda {\bf L} \cdot {\bf S}$, where $\lambda$ 
denotes the coupling strength, and ${\bf S}$ and ${\bf L}$ are spin and orbital angular momentum operators of the $t_{2g}$ orbitals, respectively. 

SOC is a relativistic effect roughly proportional to Z$^4$, where Z is the atomic number, and hence
studies so far have been limited to iridium (Z=77) and other heavy elements. Na$_2$IrO$_3$ and Li$_2$IrO$_3$, while good candidates, suffer from trigonal lattice distortions and diminished two-dimensionality (2D) due to the Na atoms sandwiched between the honeycomb layers. The correct low-energy description is also under debate: single SOC-induced $j_{\rm eff}=1/2$ state versus nonrelativistic molecular orbitals.\cite{mazin_prl,hskim_prb} 
Thus, the search for more ideal 2D honeycomb materials described by a $j_{\rm eff}=1/2$ picture is important.

Recently, it was suggested that a ruthenium chloride $\alpha$-RuCl$_3$ (RuCl$_3$) is a good candidate  because of its more ideal 2D honeycomb structure.\cite{plumb2014alpha} Although RuCl$_3$ should be metallic given the partially filled bands from the five valence electrons in $t_{\rm 2g}$ orbitals,
an insulating behavior is observed\cite{binotto1971optical,pollini1996electronic}, suggesting the possibility of a Mott insulating phase driven by electron 
correlations. A natural question follows about the role of SOC; naively one would expect that it would not play a major part as atomic SOC in Ru is $\lambda \sim 0.1$~eV\cite{porterfield2013inorganic}, a fraction of that in Ir.



\begin{figure}[t]
	\centering
	\includegraphics[width=0.45 \textwidth]{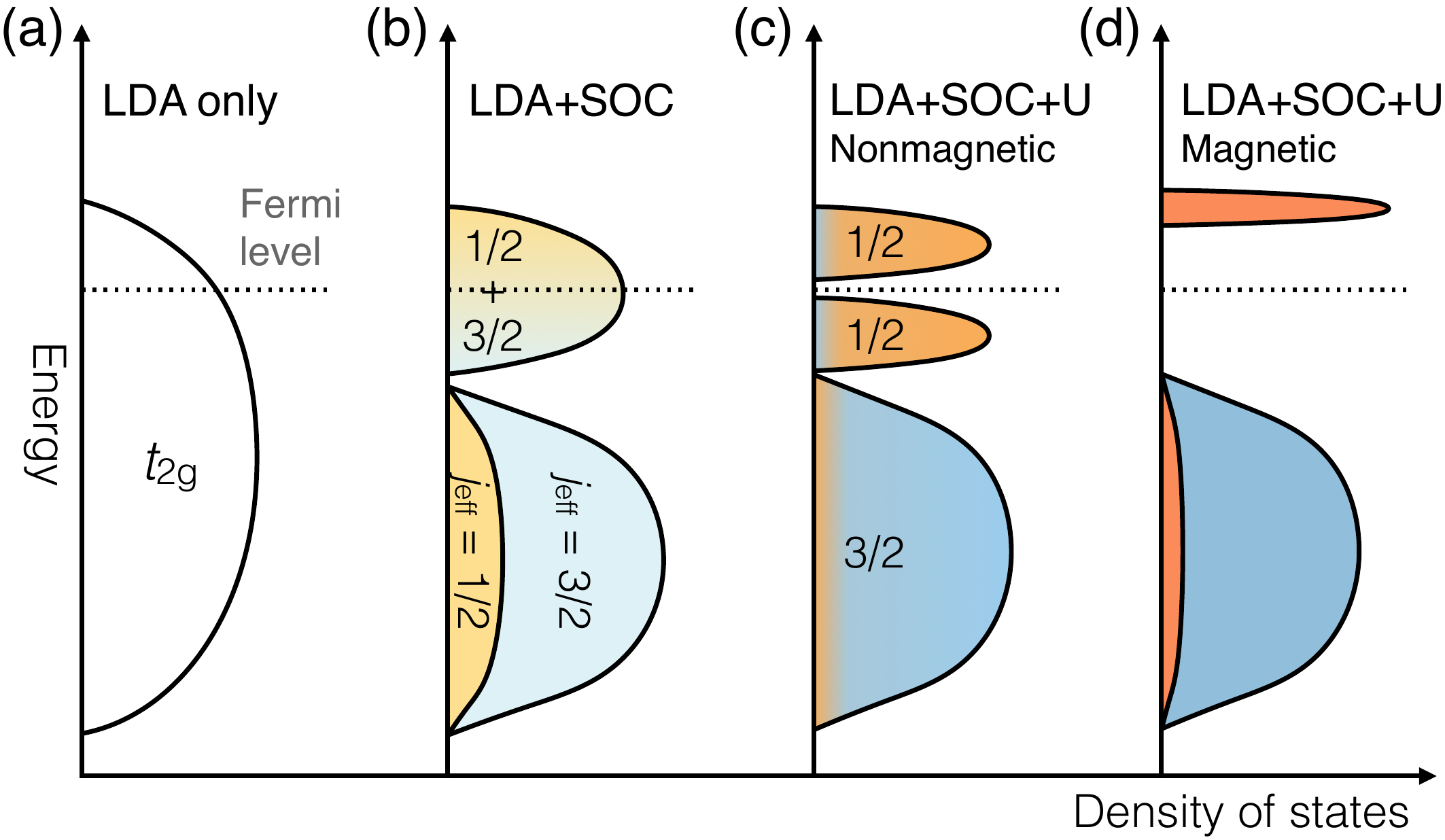}
	\caption{(Color online)
          Schematic diagrams depicting the density of states (DOS) and change in the electronic structure of RuCl$_3$
          as SOC and the on-site Coulomb interactions are included. Red and blue colors represent the weights of $j_{\rm eff}$ = 1/2 and 3/2 states, respectively. Panel (a) displays the DOS without SOC, and panel (b) shows the DOS with  SOC, which shows no clear separation between $j_{\rm eff}$ =1/2 and 3/2 bands. On including U and fixing a paramagnetic state, as shown in panel (c), the  bands near the Fermi level acquire $j_{\rm eff}$ = 1/2 character and are separated from the 3/2 bands. Panel (d) is the DOS in a magnetic ground state realized in RuCl$_3$.  
        }
	\label{fig:scheme}
\end{figure}

\begin{figure*}
  \centering
  \includegraphics[width=1.0 \textwidth]{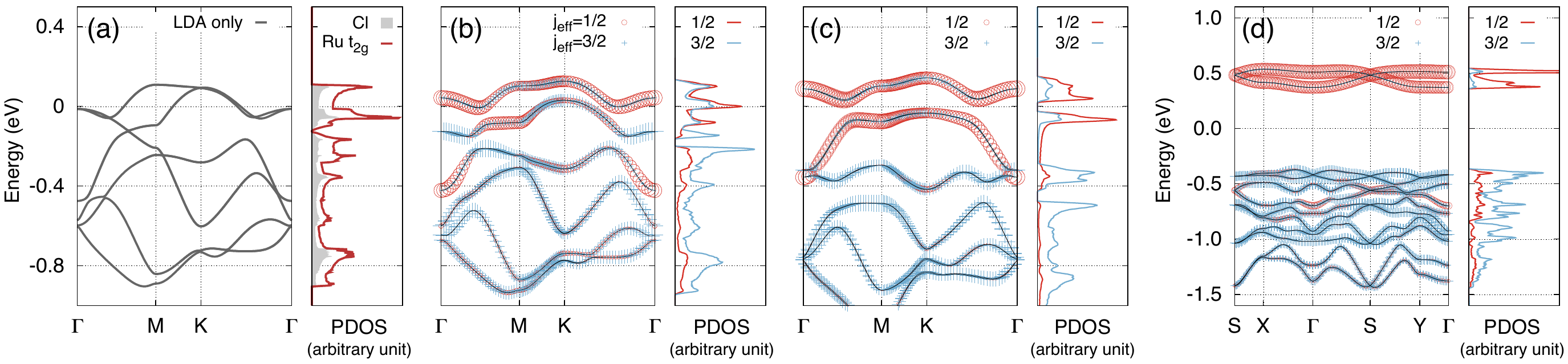}
  \caption{(Color online)
          (a) Electronic structure of RuCl$_3$ without SOC and electron interactions.
          Red and gray curves depict the projected density of states (PDOS)
         for Cl, and Ru $t_{\rm 2g}$, respectively.
         The $j_{\rm eff}$-projected band structures and density of states are laid out in the presence of SOC in (b), 
          SOC and the on-site Coulomb interaction of $U_{\rm eff}$ = 1.5~eV while fixing a non-magnetic state in (c), 
          and with the lowest energy zigzag (ZZ) magnetic order in (d), respectively.  
       }        
  \label{fig:bandswomag}
\end{figure*}

In this letter, we demonstrate that Kitaev magnetism can indeed be achieved in RuCl$_3$ despite its smaller atomic SOC strength.
We arrive at this conclusion by first studying the role of electronic correlations using {\it ab-initio} electronic structure calculations.
The results are summarized in the schematic density of states (DOS) depicted in Fig. 1.
The $t_{\rm 2g}$ bands without SOC are shown in Fig. 1(a).
In the presence of SOC, the bands near the Fermi level are mixtures of $j_{\rm eff}$ = 1/2 and 3/2 shown in Fig. 1(b).
This mixing is quite a contrast to the band structure of iridates, where the $j_{\rm eff}$ = 1/2 and 3/2 bands are well separated. 
Nevertheless, when the on-site Coulomb interaction $U$ is introduced while fixing a paramagnetic state, the bands near the Fermi level take on a predominantly $j_{\rm eff}=1/2$ character and a band gap develops as shown in Fig. 1 (c), suggesting a correlation induced insulating phase. 
We further derive a spin Hamiltonian and determine spin exchange parameters in the strong SOC limit employing tight binding parameters obtained by projecting the {\it ab-initio} band structure. We find that zigzag (ZZ) magnetic order has the lowest energy, and its corresponding band structure is shown in Fig. 1 (d). We also discuss experimental tools to test our theory.

{\it Ab-initio calculations} --
RuCl$_3$ has a layered honeycomb structure and a $d^5$ valence electron configuration for Ru$^{3+}$, similar to the Ir$^{4+}$ ion in Na$_2$IrO$_3$. While Na$_2$IrO$_3$ suffers from considerable lattice distortions,  RuCl$_3$ has nearly perfect local cubic symmetry. Since the honeycomb layers of RuCl$_3$ are weakly coupled, we study a single honeycomb layer which should capture the important physics. We used OpenMX\cite{openmx,*openmx2}, which employs linear-combination-of-pseudo-atomic-orbitals, for the electronic structure calculations and confirmed our results with the Vienna Ab initio Simulation Package\cite{VASP1,VASP2}. Further details about our calculations are in the Supplementary Material\footnote{See Supplemental Material for computational details and further information on tight-binding parameters and exchange interactions.}.

The results of electronic structure calculations are presented in Fig. \ref{fig:bandswomag}. 
Fig. \ref{fig:bandswomag}(a) shows the bands and projected density of states (PDOS) of RuCl$_3$ 
without SOC and electronic interactions. 
The long Ru-Cl and Ru-Ru bonds
result in a Ru $t_{\rm 2g}$ bandwidth of only 1~eV, significantly smaller than the bandwidth of honeycomb iridates\cite{mazin_prl,foyevtsova2013ab,chkim_prl,hskim_prb}.
The smaller bandwidth of RuCl$_3$ makes it more susceptible to SOC and correlations compared to its 5$d$ counterparts. 
On the other hand, since each band in the $t_{\rm 2g}$ manifold disperses across the entire bandwidth, the quasi-molecular orbital picture suggested for Na$_2$IrO$_3$ is unsuitable for RuCl$_3$\cite{mazin_prl}. Further clarification is provided in the supplementary material where the overlaps between the $t_{\rm 2g}$ orbitals obtained by the maximally-localized Wannier orbital method\cite{marzari1997maximally} is described.

In the presence of SOC, the band structure and PDOS projected onto the $j_{\rm eff}$ states are shown in Fig. \ref{fig:bandswomag}(b). 
The magnitude of Ru SOC is found to be 0.14~eV, which is small compared to the bandwidth. 
While one can distinguish the $j_{\rm eff}$ = 1/2 and 3/2 bands near the $\Gamma$-point, 
they are mixed with each other near the Brillouin zone boundaries, especially near the K-point. 
PDOS shows that the $j_{\rm eff}$-projected weights of the 1/2 and 3/2 states near the Fermi level are comparable, showing that unlike its 5$d$ counterpart Na$_2$IrO$_3$, SOC alone is insufficient to support
the $j_{\rm eff}$ = 1/2 picture in RuCl$_3$. The on-site Coulomb interactions in Ru $d$-orbitals, however, 
does promote the $j_{\rm eff}$ = 1/2 picture.

We performed LDA+SOC+$U$ calculations fixing a paramagnetic (PM) phase to understand the combined effects of interactions and SOC without a magnetic order.
Fig. \ref{fig:bandswomag}(c) shows the PM results with $U_{\rm eff} \equiv U - J_{\rm H}$ = 1.5~eV ($J_{\rm H}$ is Hund's coupling), 
which is a metastable solution that can be obtained by slowly increasing $U_{\rm eff}$ from the noninteracting starting point.
Compared to Fig. \ref{fig:bandswomag}(b), one can see that the $j_{\rm eff}$ = 3/2 states are pushed down significantly,  
so that the low-energy states near the Fermi level can be described purely in terms of the $j_{\rm eff}$ = 1/2 states. 
The effective SOC at $U_{\rm eff}$ = 1.5~eV is about twice the atomic value, a dramatic 
enhancement compared to results reported for iridates recently\cite{hck_arxiv}. Previously, such an enhancement was reported for the 4$d$ transition metal oxide Sr$_2$RhO$_4$.\cite{Liu_SRO}

\begin{figure}
	\centering
	\includegraphics[width=0.4 \textwidth]{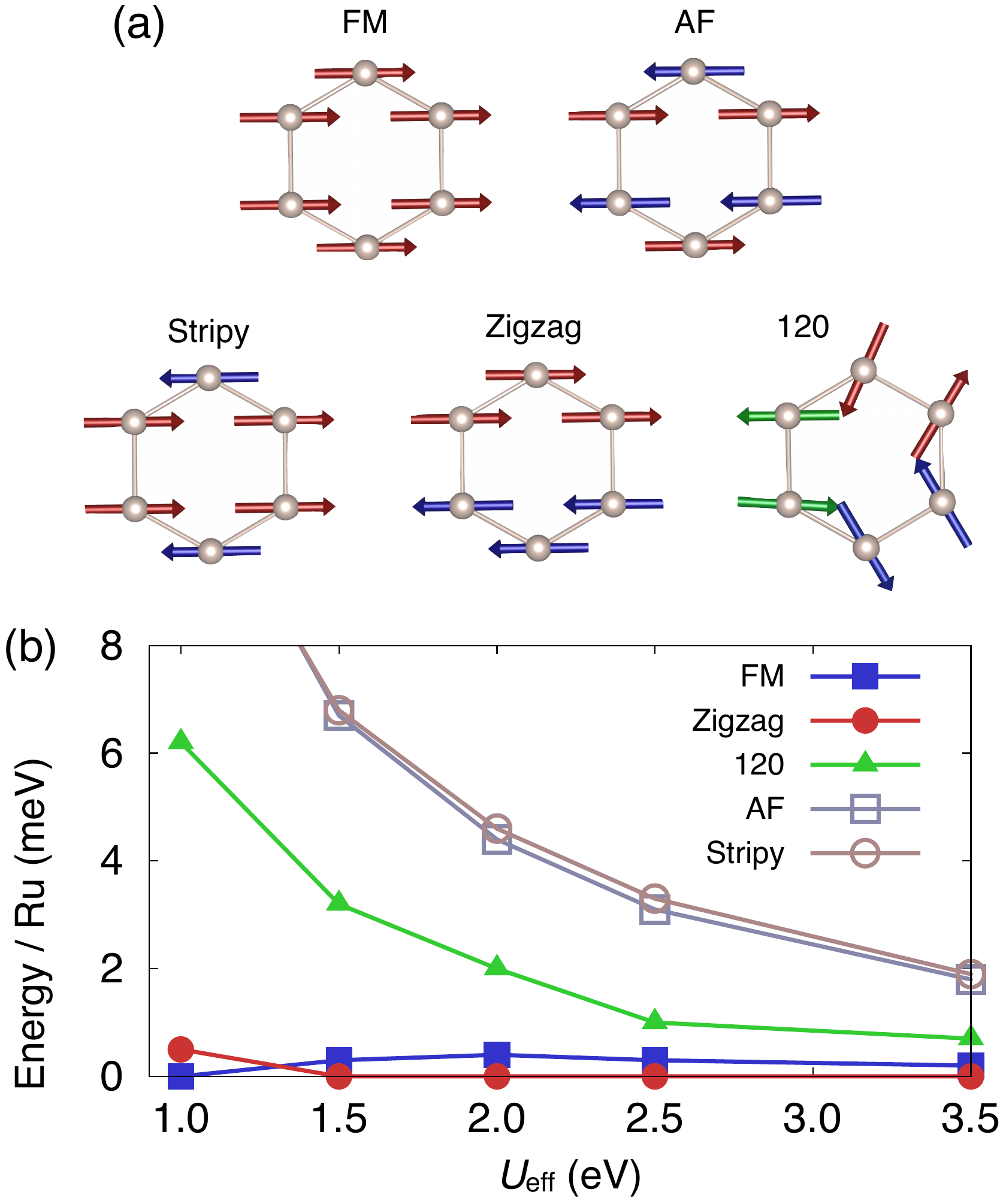}
	\caption{(Color online)
          (a) Collinear magnetic configurations considered in the LDA+SOC+$U$ calculations. 
          (b) Relative energy difference per Ru atom for each configuration 
          plotted with respect to $U_{\rm eff}$. The ZZ ordered state has the lowest energy except when $U_{\rm eff}$ =1.0eV, but FM is competitive and the 120 ordered state approaches both states in energy when $U_{\rm eff}$ is large.
        }
	\label{fig:magldau}
\end{figure}

Having established how correlations lead to a $j_{\rm eff}$ = 1/2 picture in RuCl$_3$, we studied the energies of five different magnetic phases shown in Fig. \ref{fig:magldau} (a) -- ferromagnet (FM), antiferromagnet (AF), stripy (ST), zigzag (ZZ), and 120 order.
The relative energy differences between these phases as a function of $U_{\rm eff}$ is shown in Fig. \ref{fig:magldau} (b). 
We find that the ZZ phase is the ground state over the entire range of $U_{\rm eff}$ up to 3.5~eV, except at $U_{\rm eff}$ = 1.0~eV where the FM phase has lower energy. In the higher $U_{\rm eff}$ regime, ZZ is nearly degenerate with FM and 120 ordering. The electronic band structure for this ZZ state is shown in Fig. \ref{fig:bandswomag}(d).
After the magnetic order sets in, the $j_{\rm eff}$ = 1/2 bands are further pushed away (the gap increases), and the occupied $j_{\rm eff}$=1/2 band is now mixed with the $j_{\rm eff}$ = 3/2 bands.

\begin{figure}
  \centering
  \includegraphics[width=0.45 \textwidth]{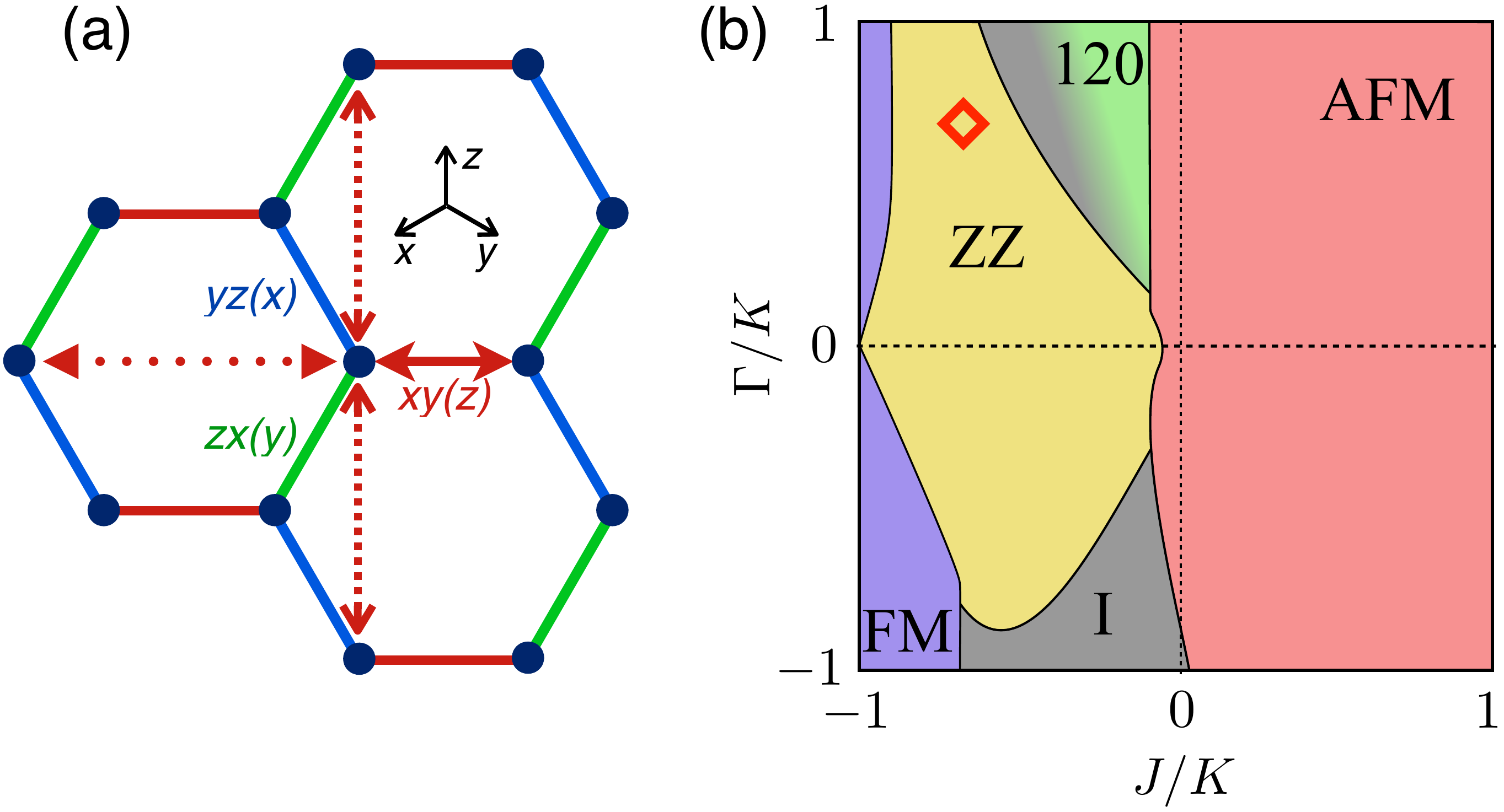}
  \caption{(Color online)
          (a) 1st (solid), 2nd (dashed), and 3rd (dotted) n.n bonds on the honeycomb lattice with the bond labels. 
          Red, blue, and green colors depict the $\alpha \beta (\gamma) = xy(z)$, $yz(x)$, and $zx(y)$ bond, respectively,	
          where $\alpha$, $\beta$, and $\gamma$ denote the spin components interacting on the specified bond.
		  Further neighbor hoppings with only $xy(z)$-type are depicted in the figure. 
          (b) shows the Luttinger-Tisza phase diagram at $J_{\rm H}/U=0.2$ for fixed 2nd and 3rd n.nexchanges. 
          Grey shading within the 120 order phase depicts the trace of incommensurate (I) order occuring in that area.
          The red diamond marks the estimated parameters for RuCl$_3$. See the main text for a description of the exchange parameters. 
        }
  \label{fig:PD}
\end{figure}

{\it $j_{\rm eff}$=1/2 spin model in the strong coupling limit} --
As RuCl$_3$ is considered a Mott insulator\cite{plumb2014alpha}, we construct a strong coupling spin model to
capture the possible magnetic phases of RuCl$_3$. Our analysis of correlation enhanced SOC allows us to
construct a spin model based on pseudospin $j_{\rm eff}$=1/2 states near the Fermi level.
On each bond we fix a spin direction $\gamma$ and label the bond $\alpha\beta(\gamma)$ as in Fig.\ref{fig:PD}(a), with $\alpha$ and $\beta$ being the remaining two spin directions. The spin Hamiltonian relevant for RuCl$_3$, obtained from {\it ab-initio} results is then,
\begin{align}
  H &= \sum_{\mathclap{\langle ij\rangle\in \alpha\beta(\gamma)}} \left(J {\bf S}_{i}\cdot{\bf S}_{j} + K S^\gamma_i S^\gamma_j + \Gamma(S^\alpha_iS^\beta_j + S^\beta_i S^\alpha_j) \right) \nonumber \\
  &+ \sum_{\mathclap{\langle \langle  ij\rangle \rangle \in \alpha\beta(\gamma) }}\left(J_2^\alpha S^\alpha_i S^\alpha_j + J_2^\beta S^\beta_i S^\beta_j + J_2^\gamma S^\gamma_i S^\gamma_j\right) \\
  &+ \sum_{\mathclap{\langle \langle \langle ij\rangle \rangle \rangle \in \alpha\beta(\gamma)}} \left(J_3 {\bf S}_i \cdot {\bf S}_j + K_3 S_i^\gamma S_j^\gamma  +\Gamma_3(S^\alpha_iS^\beta_j + S^\beta_iS^\alpha_j)\right), \nonumber
\end{align}
where $i, j$ label the Ru$^{3+}$ sites and ${\bf{S}}_i$ is a $j_{\rm eff}$=1/2 spin operator with components $S^\alpha_i$. The parameters $J$ and $K$ are Heisenberg and Kitaev exchanges respectively, and $\Gamma$ is a symmetric off-diagonal exchange. $J_2^{(x,y,z)}$ are anisotropic spin exchanges at the 2nd n.n level while $J_3$, $K_3$ and $\Gamma_3$ are the 3rd n.n analogues to the n.n exchanges. 


Since the exchanges are expressed in terms of overlaps between $t_{\rm 2g}$ states, the on-site Coulomb interaction $U$, and the Hund's coupling $J_{\rm H}$, they can be estimated using the tight-binding parameters deduced from the {\it ab-initio} calculations. For fixed $J_{\rm H}/U = 0.2$, we find that the n.n terms dominate with antiferromagnetic $K$, ferromagnetic $J$, and positive $\Gamma$. 
Including n.n $t_{\rm 2g}$-$e_{\rm g}$exchange processes in addition to the ones within $t_{\rm 2g}$, we estimate the n.n exchanges to be
$J/K \simeq -0.7$ and $\Gamma/K \simeq 0.7$.
 The estimates for the 2nd n.n exchanges on a $z$-bond denoted by red dashed lines in Fig.\ref{fig:PD} are $J_2^x/K \simeq -0.03$, $J_2^y/K \simeq -0.01$, $J_2^z/K \simeq -0.01$ and those for 3rd n.n are $J_3/K \simeq 0.02$, $K_3/K \simeq 0.03$ with vanishingly small $\Gamma_3/K$.
We note that the Kitaev exchange is further enhanced due to inter-orbital $t_{\rm 2g}$-$e_{\rm g}$ hopping.\cite{chaloupka2013zigzag} For more details, including explicit expressions for the exchanges and tight-binding parameters, 
see the Supplementary Material.

Luttinger-Tisza analyses\cite{LuttingerTisza} were performed to obtain classical ground states of the above model. 
A phase diagram for varying $J/K$ and $\Gamma/K$ while keeping $J^{(x,y,z)}_2/K$, $J_3/K$, and $K_3/K$ fixed is presented in Fig. 4(b). Based on the strength of the exchanges (see Supplementary Material) we find that the relevant position for RuCl$_3$, denoted by a red diamond in Fig. 4(b), is in 
the ZZ regime close to FM and 120 ordered states. While the qualitative features of the phase diagram are well captured by the n.n $J$-$K$-$\Gamma$ model, addition of 2nd and 3rd n.n exchanges enlarges the ZZ region. This enhancement of the ZZ phase on adding further neighbor exchanges was also observed for $J_H/U = 0.3$ and is likely
independent of the $J_H/U$ ratio.
Our analysis predicts that RuCl$_3$ has a zigzag ordered ground state, described by a pseudospin $j_{\rm eff}$=1/2 model, lying close to the antiferromagnetic Kitaev spin liquid. It is remarkable that the ZZ phase 
is surrounded by FM and 120 ordered phases in the strong-coupling phase diagram, these states are also found to be very close in energy in our LDA+SOC+$U$ calculations.




{\it Discussion and Conclusion} --
There are various experimental ways to test our proposal.
One experimental technique is angle resolved photoemission spectroscopy (ARPES), which is ideal for RuCl$_3$ with its layered structure. Occupied states below the Fermi level
should reflect a large gap as well as flat dispersion across the Brillouin zone. 

In the iridates, the first measurement that stimulated the idea of Sr$_2$IrO$_4$ being a spin-orbit Mott insulator was the optical conductivity, where an optical gap of around 0.5~eV was seen\cite{kim2008novel}. In RuCl$_3$ however, previous optical data was interpreted in terms of a small optical gap of 0.2 - 0.3~eV, but the extremely small intensity in this region suggests that this feature may not be associated with charge excitations\cite{luke}.
Provided the optical gap is identified with the onset of the peak at around 1eV in existing studies\cite{plumb2014alpha,binotto1971optical,guizzetti1979optical},  which is bigger than the observed values of 0.5~eV in Sr$_2$IrO$_4$\cite{kim2008novel} and 0.34~eV in Na$_2$IrO$_3$\cite{comin2012Na}, our results are in good agreement with the optical data.

Our prediction of ZZ magnetic order in the ground state, should be detectable by neutron scattering. An elastic neutron scattering measurement that has just been reported found 
a magnetic peak at the wave-vector ${\bf M}$ below 8 K\cite{jasears_arxiv}, suggesting that the magnetic order is either ZZ or ST. Based on the analysis of anisotropy in susceptibility provided in Ref. \onlinecite{rau2014generic,jasears_arxiv}, we find an antiferromagnetic $K$, a ferromagnetic $J$ which is a fraction of $K$, and a finite $\Gamma$. Thus, ZZ magnetic order should be consistent with both neutron and susceptibility data.
Inelastic neutron scattering analysis, similar to the one reported for Na$_2$IrO$_3$ \cite{choi2012spin}, can provide further confirmation, since the spin-wave spectra including spin gaps are different in the ZZ and ST phases. Thus computing spin wave excitations in various regimes of the strong-coupling model would be a natural step for a future study.

It is important to note that although RuCl$_3$ shows a ZZ ordered phase similar to Na$_2$IrO$_3$, 
the microscopic origins of the two ZZ ordered phases are quite different. 
The Kitaev interaction is antiferromagnetic in RuCl$_3$,while it is ferromagnetic in Na$_2$IrO$_3$. 
This is because the Kitaev exchange originates from oxygen mediated hopping in Na$_2$IrO$_3$,
while in RuCl$_3$, it is primarily due to direct overlap of $d$ orbitals. 
The difference between the two compounds comes from the difference of covalency between oxygen
and chlorine ions, suggesting that qualitative features of the underlying low-energy physics 
depends on structural and chemical details in these layered honeycomb compounds.
The different magnetic ground states in Na$_2$IrO$_3$ and Li$_2$IrO$_3$ which shows an incommensurate spiral magnetic order\cite{singh_Li2IrO3} is another example.
In this regard, a comparative study of RuCl$_3$ and Li$_2$RhO$_3$, which is isostructural and isoelectronic to Li$_2$IrO$_3$\cite{mazin_Li2RhO3}, can be interesting as both share similar SOC strengths and electron correlations but have different lattice constants and $p$-orbital covalency.


In summary, combining {\it ab-initio} and strong coupling approaches,
we have investigated the electronic and magnetic properties of RuCl$_3$. Our results strongly suggest that this compound
can be understood as an interaction-driven $j_{\rm eff}$ = 1/2 system, which hosts
magnetism dominated by the Kitaev interaction. Owing to the simple and ideal crystal structure, RuCl$_3$ provides an excellent platform to explore the physics of SOC and electronic correlations
as well as related unconventional magnetism. Our study also opens up the possibility of a whole new class of materials in which to explore physics driven by spin-orbit coupling and electronic correlations, beyond the 5d transition metal oxides.

{\it Acknowledgements} --
We thank Y.-J. Kim, K.S. Burch, L. Sandilands, and E.K.-H Lee for useful discussions and R. Schaffer for a critical reading of the manuscript.
This work was supported by the NSERC of Canada and the center for Quantum Materials at the University of Toronto.
Computations were mainly performed on the GPC supercomputer at the SciNet HPC Consortium. SciNet is funded by: 
The Canada Foundation for Innovation under the auspices of Compute Canada; the Government of Ontario; Ontario Research Fund for 
Research Excellence; and the University of Toronto. 
HSK thanks the IBS Center for Correlated Electron System at Seoul National 
University for additional computational resources and and V.S.V thanks NSERC-CREATE for a graduate fellowship through the HEATER program. 

\bibliography{short-titles,rucl3}

\pagebreak
\widetext


\setcounter{equation}{0}
\setcounter{figure}{0}
\setcounter{table}{0}
\setcounter{page}{1}
\makeatletter
\renewcommand{\theequation}{S\arabic{equation}}
\renewcommand{\thefigure}{S\arabic{figure}}
\renewcommand{\bibnumfmt}[1]{[S#1]}

\section{\label{app:dft_details}Supplementary Material A:
Details on $t_{\rm 2g}$ overlaps }

\begin{figure}[tb]
  \centering
  \includegraphics[width=0.95 \textwidth]{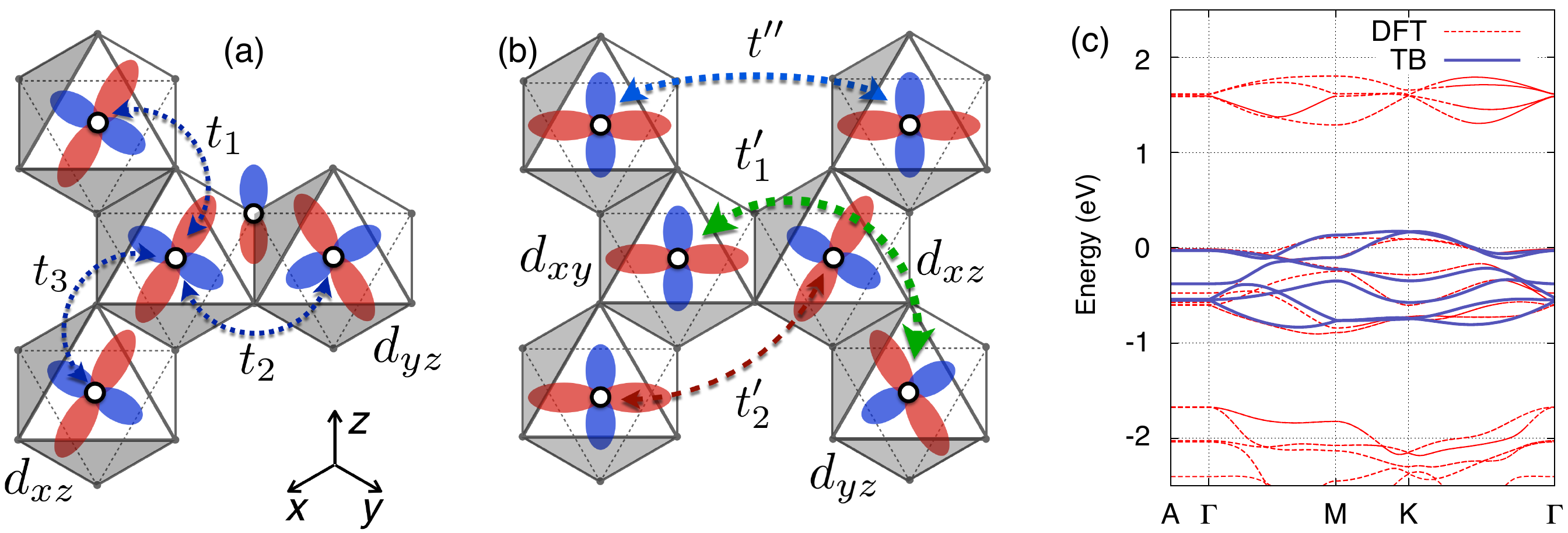}
  \caption{(Color online) 
  (a-b) Six major hopping channels between neighboring Ru $t_{\rm 2g}$ orbitals,
  where (a) shows the nearest-neighbor (n.n) and (b) shows second (2nd) and third (3rd) 
  n.n overlaps .
  (c) Band structure of RuCl$_3$ monolayer, where dashed red lines show {\it ab-initio} bands without SOC and on-site Coulomb interactions, and
  the tight-binding (TB) bands from the overlaps depicted in 
  (a) and (b) are in solid blue.
  }
  \label{figA:NNNN}
\end{figure}

In order to understand the hopping processes between Ru $t_{\rm 2g}$ orbitals better and for estimating the exchange interactions, we calculated the overlaps using maximally-localized Wannier orbital 
calculations without and with the presence of SOC\cite{marzari1997maximally,weng2009revisiting}. Due to the 
virtually ideal crystal structure and large distances between Ru atoms,
six major hoppings suffice to reproduce the electronic structure. Since the inclusion of SOC induce only small change (less than 2 meV) to the hopping terms except the on-site SOC terms, here we show the results from non-relativistic calcualtions.
Fig. \ref{figA:NNNN}(a) shows the three nearest neighbor (n.n) 
hoppings; $t_1$, $t_2$, and $t_3$, where $t_1$ and $t_3$ are intra-orbital overlaps where as $t_2$ is an inter-orbital overlap. The largest overlap, $t_3$, 
is dominated by the $\sigma$-type direct overlap between Ru $t_{\rm 2g}$ orbitals, while $t_2$ has contributions from $\pi$-type direct overlap in addition to indirect $d$-$p$-$d$ contribution via the $p$-orbitals of intervening Chlorine atoms. $t_1$ is due to $\pi$- and $\delta$-type direct overlap.
Small distortions in the crystal structure
give rise to anisotropies less than 5 meV in the hoppings so we take their average to obtain $t_1=65$ meV, $t_2=113$ meV, and $t_3=-226$ meV.
In the second (2nd) n.n hopping channels, depicted in Fig. \ref{figA:NNNN}(b), there are two 
inequivalent inter-orbital hoppings $t'_1$ and $t'_2$ due to the absence of inversion symmetry
with respect to the bond center. In $t'_1$ channel the orbital lobes participating in the 
hopping point towards the intermediate Ir site, while in $t'_2$ they are directed towards the center of the honeycomb. There are also third (3rd) n.n intra-orbital hoppings denoted by $t''$. 
Their magnitudes are $t'_1=-20$ meV, $t'_2=-58$ meV, and $t''=-49$ meV, and have no
direction dependence unlike in the case of n.n hopping channels.

All other overlaps are less than 10 meV and can be considered marginal;
tight-binding band structure from the six major contributions shows good agreement with the {\it ab-initio} bands in Fig. \ref{figA:NNNN}(c). 
Note that unlike in several two- and three-dimentional honeycomb iridates where $t_2$ is dominant\cite{hskim_prb,foyevtsova2013ab,HSK_HH0}, in
RuCl$_3$ $t_3$ is twice as large as $t_2$.
The quasimolecular orbital 
character in the two-dimensional iridate Na$_2$IrO$_3$\cite{mazin_prl} originates from this $t_2$ overlap. Since it is not the principal overlap in RuCl$_3$, the conventional $t_{\rm 2g}$ orbital picture 
is more appropriate for understanding the electronic structure. Detailed hopping magnitudes are provided in Table \ref{tabA:hops}.

\begin{table}[tb!]
  \centering
  \begin{tabularx}{\textwidth}{lXrrrXrrrXrrr} \hline\hline 
      && \multicolumn{3}{r}{n.n} && \multicolumn{3}{r}{2nd n.n} && \multicolumn{3}{r}{3rd n.n} \\
      && \multicolumn{3}{r}{${\bf r}_{ij} = (-d, +d, 0)$} 
        && \multicolumn{3}{r}{${\bf r}_{ij} = (-d, -d, +2d)$} 
        && \multicolumn{3}{r}{${\bf r}_{ij} = (-2d, +2d, 0)$} \\
      && \multicolumn{3}{r}{~~~~~~~~~~~~~~~~~~~~$A \rightarrow B$} 
        && \multicolumn{3}{r}{~~~~~~~~~~~~~~~~~~~~$A \rightarrow A$} 
        && \multicolumn{3}{r}{~~~~~~~~~~~~~~~~~~~~$B \rightarrow A$} \\ \hline
      ${\bf T}_{ij}$  && $d_{xy}$ & $d_{yz}$ & $d_{xz}$ && $d_{xy}$ & $d_{yz}$ & $d_{xz}$ && $d_{xy}$ & $d_{yz}$ & $d_{xz}$ \\
      $d_{xy}$ && -0.229   & -0.010   & -0.011   &&  0.000   & +0.006   & +0.004   && -0.049   & +0.009   & +0.009 \\
      $d_{yz}$ && -0.011   & +0.065   & +0.114   && +0.003   &  0.000   & -0.020   && +0.010   & -0.008   & -0.005 \\
      $d_{xz}$ && -0.009   & +0.113   & +0.066   && +0.006   & -0.058   &  0.000   && +0.009   & -0.005   & -0.008 \\ \hline\hline
  \end{tabularx}
  \caption{(Unit in eV) A subset of Ru $t_{\rm 2g}$ hoppings ${\bf T}_{ij}$ as representatives 
  of each hopping channel up to third NN, 
  where $H_{t_{\rm 2g}} = \sum_{ij} {\bf C}^{\dag}_{i} \cdot {\bf T}_{ij} \cdot {\bf C}_j$
  with ${\bf C}^{\dag}$ and ${\bf C}$ being the creation and annihilation operators for
  $t_{\rm 2g}$ states, respectively. $A$ and $B$ are sublattice indices, and
  ${\bf r}_{ij}$ is expressed in terms of the coordinates
  depicted in Fig. \ref{figA:NNNN}(a-b), where
  $d\simeq2.43$\AA~ is approximate distance between Ru and Cl, 
  Other hoppings  can be recovered by applying ${\bf T}_{ji} = {\bf T}^{\dag}_{ij}$,
  ${\bf T}_{A\rightarrow A} = ({\bf T}_{B\rightarrow B})^\dag$,
  $C_{3}$ rotations along the threefold axis perpendicular to the honeycomb plane, 
  and inversion operations.
  }
  \label{tabA:hops}
\end{table}

\section{\label{app:hops}Supplementary Material B:
Exchange interactions}

The pseudospin $j_{\rm eff}=1/2$ spin Hamiltonian (Eq.1 in the main text) is,
\begin{align*}
  H &= \sum_{\langle ij\rangle\in \alpha\beta(\gamma)} \left(J {\bf S}_{i}\cdot{\bf S}_{j} + K S^\gamma_i S^\gamma_j + \Gamma(S^\alpha_iS^\beta_j + S^\beta_i S^\alpha_j) \right) \\
  &+ \sum_{\langle\langle ij\rangle\rangle \in \alpha\beta(\gamma) } \left(J_2^\alpha S^\alpha_i S^\alpha_j + J_2^\beta S^\beta_i S^\beta_j + J_2^\gamma S^\gamma_i S^\gamma_j\right) \\
  &+ \sum_{\langle\langle\langle ij\rangle\rangle\rangle\in \alpha\beta(\gamma)} \left(J_3 {\bf S}_i \cdot {\bf S}_j + K_3 S_i^\gamma S_j^\gamma + \Gamma_3(S^\alpha_iS^\beta_j + S^\beta_iS^\alpha_j)\right).
\end{align*}
Explicit expressions of the exchange interactions for the 
n.n $j_{\rm eff}$=1/2 spins derived in Ref. \onlinecite{rau2014generic} are:
\begin{align}
\label{eq:nn}
J &= \frac{4}{27}\left[ 
\frac{6t_1 (t_1+2t_3)}{U-3J_{\rm H}} + 
\frac{2(t_1-t_3)^2}{U-J_{\rm H}} + \frac{(2t_1 + t_3)^2}{U+2J_{\rm H}} 
\right] \nonumber \\
K &= \frac{8J_{\rm H}}{9}\left[ 
\frac{(t_1-t_3)^2 - 3t^2_2}{(U-3J_{\rm H})(U-J_{\rm H})} 
\right] \nonumber \\
\Gamma &= \frac{16J_{\rm H}}{9}\left[ 
\frac{t_2(t_1-t_3)}{(U-3J_{\rm H})(U-J_{\rm H})}
\right]. 
\end{align}
For the 2nd n.n interactions, due to the asymmetry of the hopping
channels al the three $J^\alpha_2~(\alpha=x,y,z)$ exchanges are different. In terms of
$t'_1$, $t'_2$, $U$ and $J_{\rm H}$ they can be written as below:
\begin{small}
\begin{align}
\label{eq:nnn2}
J^x_2 &=
-\frac{4U}{9} \frac{(t'_1 - t'_2)^2}{(U+2J_{\rm H})(U-3J_{\rm H})} + 
\frac{4J_{\rm H}}{9} \frac{(t'^2_2 - t'^2_1 )}{(U-J_{\rm H})(U-3J_{\rm H})(U+2J_{\rm H})}
\simeq -J^0_2 + \frac{8}{9U} t'(t'-t'_p)   \left( \frac{J_{\rm H}}{U} \right), \nonumber \\
J^y_2 &=
-\frac{4U}{9} \frac{(t'_1 - t'_2)^2}{(U+2J_{\rm H})(U-3J_{\rm H})} - 
\frac{4J_{\rm H}}{9} \frac{(t'^2_2 - t'^2_1 )}{(U-J_{\rm H})(U-3J_{\rm H})(U+2J_{\rm H})}
\simeq -J^0_2 + \frac{8}{9U} t'_p(t'_p-t') \left( \frac{J_{\rm H}}{U} \right), \nonumber \\
J^z_2 &=   
+\frac{4U}{9} \frac{(t'_1 - t'_2)^2}{(U+2J_{\rm H})(U-3J_{\rm H})} - 
\frac{4J_{\rm H}}{9} \frac{(t'_1 + t'_2)^2 + 2t'_1 t'_2}{(U-J_{\rm H})(U-3J_{\rm H})(U+2J_{\rm H})}
\simeq +J^0_2 - \frac{8}{9U} (t'+t'_p)^2   \left( \frac{J_{\rm H}}{U} \right),
\end{align}
\end{small}
for the $xy(z)$ bond, where $J^0_2=4 \frac{(t'_1-t'_2)^2}{9U}$.
Note that, when the Hund's coupling becomes small,  the 2nd n.n interactions reduce to 
ferromagnetic Heisenberg ($\simeq$$-J^0_2$) and antiferromagnetic Kitaev ($\simeq$$2J^0_2$)
interactions with the Kitaev exchange having twice the magnitude of the Heisenberg 
exchange. Similar 2nd n.n exchanges have also been observed for other layered honeycomb iridates\cite{hskim_prb,Reuther}. 
Lastly, the expression for the 3rd n.n exchange interactions are obtained by substituting 
$t_3$ to $t''$ and setting $t_1$ and $t_2$ to be zero in Eq. (1). 

Additional superexchange process through the $e_{\rm g}$-excited states
is possible and can be non-negligible due to the small energy splitting of $\Delta E_{e_{\rm g} - t_{\rm 2g}}\sim1.5$~eV between them as shown in Fig. \ref{figA:NNNN}(c). 
In Ref. \onlinecite{Khaliullin_eg} it was suggested that, this $t_{\rm 2g}$-$e_{\rm g}$ process
can give rise to ferromagnetic Heisenberg and antiferomagnetic Kitaev 
interactions \footnote{Among the four processes mentioned in Ref. \cite{Khaliullin_eg} ---
i) $\sigma$-type direct hopping, ii) $t_{\rm 2g}$-$e_{\rm g}$ process, iii) $d$-$p$-$d$
indirect process, and iv) $p$-$d$ charge transfer excitation ---, i) and iii) are already 
taken into account in our results, and iv) should be less significant compared to the other three contributions due to the weaker hybridization between the Ru $t_{\rm 2g}$ and Cl $p$ orbitals as shown
in Fig. 2(a) in the main text.}. 
Owing to the nearly distortion-free structure of RuCl$_3$, only one hopping channel 
is active in this process; $d_{xy}$-to-$d_{3z^2-r^2}$ hopping along the horizontal ``$z$-bond'' 
in Fig. \ref{figA:NNNN}(a) and (b). 
The expression is, as shown in Ref. \onlinecite{Khaliullin_eg}, as follows;
\begin{align}
H' &= \sum_{\langle ij \rangle \in \gamma} 
  I'(2S^\gamma_i S^\gamma_j - {\bf S}_i \cdot {\bf S}_j) \nonumber \\
I' &\simeq \frac{4}{9\tilde{U}}\tilde{t}^2\left( \frac{\tilde{J}_{\rm H}}{\tilde{U}} \right),
\label{eq:eg}
\end{align}
where $\tilde{J}_{\rm H} \sim J_{\rm H}$ and $\tilde{U}$ are the Hund's coupling and the 
effective excitation energy between the $t_{\rm 2g}$ and $e_{\rm g}$ states, 
and $\tilde{t}=190$~meV is the
$t_{\rm 2g}$-$e_{\rm g}$ overlap obtained from the Wannier orbital calculations. 
The two processes --- intra-$t_{\rm 2g}$ and 
$t_{\rm 2g}$-$e_{\rm g}$ processes --- have the same direction dependence, 
and both the Kitaev interactions add up to yield a larger one.
Using $J_{\rm H}/U=0.2$, $U \approx 3$eV
and $\tilde{U} \approx 1.5$eV, we get the ratios
$J/K \simeq -0.7$ and $\Gamma/K \simeq 0.7$,
$J_2^x/K \simeq -0.03$, $J_2^y/K \simeq -0.01$, $J_2^z/K \simeq -0.01$,
$J_3/K \simeq 0.02$, $K_3/K \simeq 0.03$, and vanishingly small $\Gamma_3/K$ as 
presented in the main text.

\section{\label{app:comp_details}Supplementary Material C: Computational details}

For the electronic structure calculations with SOC and on-site Coulomb interactions, OPENMX \cite{openmx,*openmx2}, which is 
based on the linear-combination-of-pseudo-atomic-orbitals, is used. A non-collinear DFT scheme and a fully 
relativistic $j$-dependent pseudopotential are used to treat SOC, with the Perdew-Zunger parametrization of the
local density approximation (LDA) chosen for the exchange-correlation functional\cite{perdew1981self}. 
500 Ry of energy cutoff was used for the real-space sampling, 
and $8\times8\times1$ $k$-grid was adopted for the primitive cell. On-site Coulomb interactions are treated via a simplified 
LDA+$U$ formalism implemented in OPENMX code\cite{han2006n,Dudarev}, with up to 3.5 eV of $U_{\rm eff} \equiv U-J_{\rm H}$ parameter 
used for Ru $d$-orbitals in our LDA+SOC+$U$ calculations. 
Maximally-localized Wannier orbitals method\cite{marzari1997maximally} implemented in OPENMX \cite{weng2009revisiting}, is used to obtain the tight-binding Hamiltonian for Ru $t_{\rm 2g}$ 
and $e_{\rm g}$ orbitals.

The LDA+SOC+$U$ results for the paramagnetic phase was doubled-checked using Vienna Ab-initio Simulation Package\cite{VASP1,VASP2}.
To check the effect of $J_{\rm H}$ on our results, Liechtenstein's more general LDA+$U$ formalism was employed,
which treats the role of $J_{\rm H}$ explicitly\cite{Lichtenstein}. A plane-wave energy cutoff of 400 eV and 13$\times$13$\times$1 
k-points for the k-point sampling were used.
Using $U=2.0$~eV and $J_{\rm H}=0.4$~eV, equivalent to $U_{\rm eff}=1.6$~eV, yields same results with OPENMX calculations;
effective SOC is enhanced.

\end{document}